\documentclass{article}
\usepackage[T1]{fontenc} 
\usepackage[utf8]{inputenc} 
\usepackage{ismir,amsmath,cite,url}
\usepackage{graphicx}
\usepackage{booktabs}
\usepackage{microtype}
\usepackage{paralist}
\usepackage{units}
\usepackage{amsfonts}

\usepackage{color}
\def\lbd{}	

\usepackage{lineno}

\title{AQUATK: An Audio Quality Assessment Toolkit}

\twoauthors
  {Ashvala Vinay} {Music Informatics Group \\ Georgia Institute of Technology \\ {\tt ashvala@gatech.edu}}
  {Alexander Lerch} {Music Informatics Group \\ Georgia Institute of Technology \\ {\tt alexander.lerch@gatech.edu}}

\def\authorname{Ashvala Vinay and Alexander Lerch}

\begin{document}

\maketitle

\begin{abstract}

Recent advancements in Neural Audio Synthesis (NAS) have outpaced the development of standardized evaluation methodologies and tools. To bridge this gap, we introduce AquaTk, an open-source Python library specifically designed to simplify and standardize the evaluation of NAS systems. AquaTk offers a range of audio quality metrics, including a unique Python implementation of the basic PEAQ algorithm, and operates in multiple modes to accommodate various user needs.

\end{abstract}

\section{Introduction}\label{sec:introduction}

Recent developments in Neural Audio Synthesis (NAS) are notable, but an automated, standardized evaluation framework is lacking. To address this, we introduce the Audio QUality Assessment Toolkit (AquaTk), an open-source Python library designed to be framework-agnostic. AquaTk includes well-established methods for evaluating the performance of audio synthesis models and codecs.

Traditional audio evaluation algorithms and metrics often require complex setup procedures. For example, to compute the Fréchet Audio Distance (FAD), one needs to clone the entire Google Research repository. While there is a PyTorch version, there are variations in the results depending on the deep learning framework.


AquaTk aims to resolve these challenges by offering:

\begin{compactitem}
\item An open-source toolkit for evaluating NAS algorithms,
\item A wide range of audio quality measures,
\item Testable implementations to ensure reliable, replicable outcomes,
\item Compatibility with popular deep learning frameworks such as TensorFlow, Pytorch and JAX,
\item Command line and graphical interfaces and,
\item An extensible architecture for easy addition of new metrics.
\end{compactitem}

Python's widespread use in the Music Information Retrieval (MIR) and machine learning communities makes a Python-based audio quality assessment library both useful and necessary for evaluating NAS technologies. In order to ensure replicability and reproducibility, we have written comprehensive tests where possible to ensure that each user facing function behaves appropriately.

In the following demonstration, we will cover the library's features, the algorithms it includes, and its licensing.

\section{Features of AQUATk}\label{sec:typeset_text}

The library requires the following additional packages:
\texttt{numpy}, \texttt{openl3}, \texttt{tqdm}, \texttt{panns-inference}, \texttt{jukemirlib}, \texttt{scikit-learn}, \texttt{streamlit}. 
Installation also includes PyTorch and TensorFlow as dependencies from openl3 and jukemirlib.

\subsection{Distances}

AquaTk includes the following metrics to compute difference between reference and test signals: 

\begin{compactitem}
\item Mean-squared error
\item Mean-absolute error
\item Cosine similarity
\item KL-divergence
\item SI-SDR
\item SNR
\end{compactitem}

The library also supports embedding-based distances:
\begin{compactitem}
	\item Fréchet Audio Distance \cite{kilgourFrechetAudioDistance2019} and,
	\item Kernel Distances \cite{binkowskiDemystifyingMMDGANs2018}
\end{compactitem}

Both of which are usable with both default and user-defined embeddings.

\subsection{Embedding extraction}

AquaTk supports popular embeddings such as Openl3 \cite{cramerLookListenLearn2019}, PANNSs\cite{kongPANNsLargeScalePretrained2020}, JukeMIR \cite{castellonCODIFIEDAUDIOLANGUAGE2021a}, and VGGish\cite{hersheyCNNArchitecturesLargescale2017}.

\subsection{Basic PEAQ}

AquaTk offers a unique open-source implementation of the basic Perceptual Evaluation of Audio Quality (PEAQ) algorithm \cite{thiedePEAQITUStandardfor2000}, based on the PEAQb package\footnote{\url{https://github.com/akinori-ito/peaqb-fast}}. It enables the computation of the Objective Difference Grade (ODG) using the FFT Ear Model and the following Model Output Variables (MOVs):

\textit{AvgBwRef}, \textit{AvgBwTst}, \textit{NMRtotB}, \textit{Average Distorted Block}, \textit{Maximum probability of detection}, \textit{Harmonic error over time}, \textit{RDF}, \textit{WModDiff1b}, \textit{Average modulation difference}, \textit{Average Modulation Difference with introduced modulations}, \textit{Loudness}  

PEAQ has regained attention in the realm of audio quality assessment, particularly due to its perceptually relevant features. An example is the 2F model, which uses PEAQ Model Output Variables (MOVs) to evaluate source separation performance \cite{torcoliObjectiveMeasuresPerceptual2021}. Early indications suggest that the performance of our port is comparable to the existing PEAQb package.

\section{Usage}

AquaTk can be operated in one of three modes: as a library, a command-line tool, or a browser-based app.

\begin{figure}
 \centerline{\framebox{
 \includegraphics[width=0.9\columnwidth]{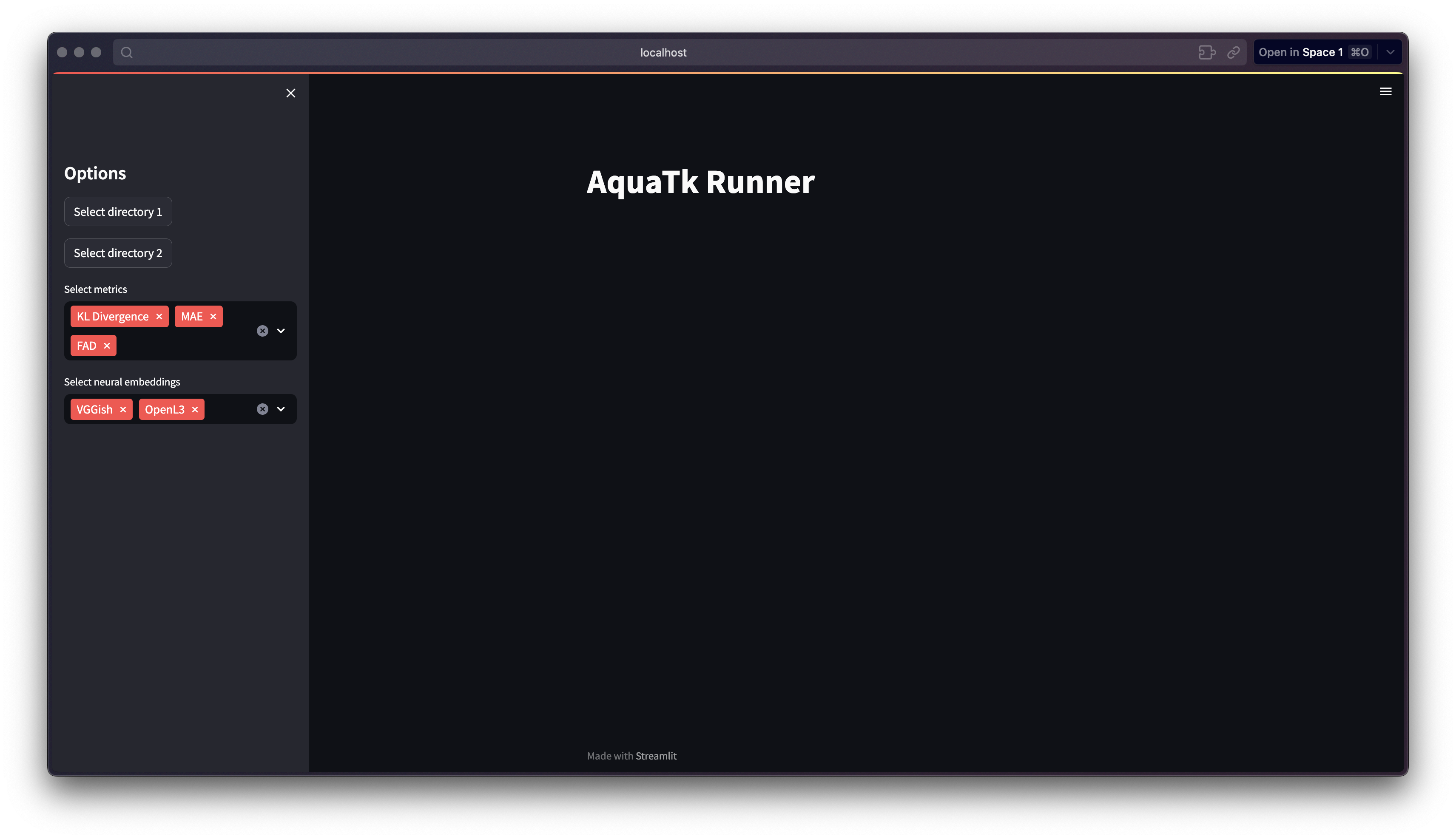}}}
 \caption{A graphical representation of AquaTK}
 \label{fig:examplegui}
\end{figure}

\begin{compactenum}
\item \textbf{Library Mode}: This mode offers flexibility by granting users direct access to the complete API and source code. Users can seamlessly integrate \texttt{aquatk} into their existing machine learning pipelines.

\item \textbf{Command-Line Mode}: Aimed at efficiency, this mode enables quick extraction of embeddings and the computation of metrics on \texttt{.wav} file directories. 

\item \textbf{Web Interface}: Constructed with Streamlit, this user-friendly mode replicates the functionalities of the command-line tool and enhances the experience with in-app visualization features. An example can be seen in \ref{fig:examplegui}.
\end{compactenum}

In Library Mode, importing \texttt{aquatk} into an existing ML pipeline grants access to the library's suite of metrics. Additionally, using AquaTk in library mode allows users to extend the package to better suit their needs.

Upon package installation, the \texttt{aquatk} command becomes available automatically. The \texttt{--web} flag launches the web interface, while various other flags can be used for command-line operations.

\section{Availability and license}

We have opted to release AquaTk under the GPL license, in accordance with the licensing terms of PeaqB. The source code will be publicly accessible via GitHub\footnote{\url{https://github.com/Ashvala/AQUA-Tk}}.


We actively encourage community involvement and contributions. One can contribute to AquaTk either by forking the repository or by submitting pull requests.

\section{Conclusion}
We present AquaTk, a Python-based, open-source library designed to address the gap in standardized evaluation methods for Neural Audio Synthesis (NAS). AquaTk supports a variety of quality metrics, including a python implementation of the basic PEAQ algorithm. The library is versatile, with three modes of operation—Library, Command-Line, and Web Interface—that cater to different user needs. Released under the GPL license, AquaTk is openly accessible and invites community collaboration.

Future work includes integrating additional metrics, improving scalability, and extending the library for real-time evaluation tasks. By providing a unified framework for evaluating NAS systems, AquaTk aims to become a valuable resource for both researchers and practitioners in the field.

\bibliography{ISMIR2023_template}

\end{document}